\documentstyle[aps,preprint]{revtex}

\def\prl#1#2#3{Phys. Rev. Lett., {\bf #1}, #2 (#3)}
\def\pla#1#2#3{Phys. Letts. A {\bf #1}, #2 (#3)}
\def\pra#1#2#3{Phys. Rev. A, {\bf #1}, #2 (#3)}
\def\pre#1#2#3{Phys. Rev. E, {\bf #1}, #2 (#3)}
\def\physd#1#2#3{Physica D {\bf #1}, #2 (#3)}

\def\beq{\begin{equation}}
\def\eqn{\end{equation}\noindent}
\begin{document}
\title{Intermittency Route to Strange Nonchaotic Attractors}
\author{Awadhesh Prasad, Vishal Mehra and Ramakrishna Ramaswamy}
\address{School of Physical Sciences, Jawaharlal Nehru University, New
Delhi 110067, India}
\date{\today}
\maketitle
\begin{abstract}
Strange nonchaotic attractors (SNA) arise in quasiperiodically driven
systems in the neighborhood of a saddle node bifurcation whereby a
strange attractor is replaced by a periodic (torus) attractor. This
transition is accompanied by Type-I intermittency.  The largest
nontrivial Lyapunov exponent $\Lambda$ is a good order--parameter for
this route from chaos to SNA to periodic motion: the signature is
distinctive and unlike that for other routes to SNA.  In particular,
$\Lambda$ changes sharply at the SNA to torus transition, as does the
distribution of finite--time or $N$--step Lyapunov exponents,
P($\Lambda_N$).
\end{abstract}
Strange nonchaotic attractors (SNAs), which are commonly found in
quasiperiodically forced systems, are geometrically strange---they are
properly described by a fractal dimension---but the largest nontrivial
Lyapunov exponent $\Lambda$ is negative, implying nonchaotic dynamics.

Since they were first described \cite{GOPY}, and shown to be generic in
quasiperiodically driven nonlinear systems \cite{general}, a number of
characteristics of SNAs have been studied, including the important
question of how they are created. Three main mechanisms or scenarios for
the creation of SNAs have been advanced. Heagy and Hammel
\cite{HH} identified the birth of a SNA with the collision between a
period-doubled torus and its unstable parent. This mechanism is
analogous to the attractor--merging crisis that occurs in chaotic
systems \cite{GORY}.  Kaneko \cite{K,NK} has described the
``fractalization'' of a torus, namely the increasing wrinkling of tori
which leads to the appearance of a SNA without any interaction with a
nearby unstable periodic orbit.  This route to SNA (and eventually to
chaos) has also been observed in several systems.  Yal\c{c}inkaya and
Lai \cite{YL} have recently shown a third route to SNAs via a blowout
bifurcation \cite{blowout}, namely through the loss of transverse
stability of a torus.

We describe a new mechanism for the creation of SNAs.  As a function of
driving parameter, a strange attractor disappears and is eventually
replaced by a 1-frequency torus through an analogue of the saddle-node
bifurcation. In the vicinity of this crisis--like phenomenon
\cite{GORY} the attractor is strange and nonchaotic. We show that the
dynamics at this transition is intermittent, and the scaling behaviour
is characteristic of Type I intermittency \cite{PM}. Furthermore, the
signature of the transition in $\Lambda$ (or in the distribution of
finite--time Lyapunov exponents) is distinctive and very different from
the routes to SNA that have been hitherto discussed. The present
mechanism is general and is likely to be operative in any
quasiperiodically driven system.

For definiteness, we illustrate our results using the quasiperiodically
forced logistic map~\cite{HH} 
\begin{eqnarray}
\label{HH}
x_{n+1} &=& \alpha (1 + \epsilon \cos (2 \pi \phi_n))~x_n~ ( 1 - x_n)
\nonumber \\ 
\phi_{n+1} &=& \phi_n + \omega ~~~~~({\rm mod ~} 1).
\end{eqnarray}\noindent
where we take $\omega = (\sqrt{5}-1)/2$, the golden mean, and $2 \le
\alpha \le 4$.  This system is convenient to study since the
phenomenology is smoothly related to that of the logistic map in the
limit of $\epsilon \to 0$. Since the driving term is multiplicative, it
is clear that the motion will remain bound if $\alpha (1 + \epsilon
\cos (2 \pi \phi_n)) \in [0,4]$, and thus for any $\alpha$, the largest
value of $\epsilon$ allowed is $4/\alpha -1$. It is preferable to work
with the rescaled driving parameter, $\epsilon^{\prime} =
\epsilon/(4/\alpha -1 )$ since this scales the region of interest in
parameter space to $0 \le \epsilon^{\prime} \le 1$. 

Fig.~1 is a phase--diagram of the system showing the different
possible dynamical behaviours---periodic, strange nonchaotic and
chaotic attractors, corresponding to the symbols P, S and C. The
different phases are characterised through the largest 
nonzero Lyapunov exponent $\Lambda$~$= \lim_{N \to \infty}
\frac{1}{N}\sum_{i=1}^N \ln \vert (1-2 x_i) \alpha(1+\epsilon \cos(2\pi
\phi_i))\vert$, which we calculate typically from 10$^6$ iterations of
the map, as well as the phase--sensitivity exponent \cite{PF}. The
logistic map with additive driving \cite{NK} has identical behaviour.

SNAs occur along the boundary of the chaotic regions \cite{SFKP}, when
$\Lambda$ takes small negative values. In the
$\alpha$--$\epsilon^{\prime}$ plane, the regions of P, S and C
behaviour can be interwoven. In particular, there are two distinct
chaotic phases: the regions marked C$_1$ and C$_2$ which are separated
by a narrow tongue where the motion is periodic.  The behaviour of
$\Lambda$ as a function of $\alpha$ for $\epsilon^{\prime} = 1 $,
namely along the upper edge of Fig.~1, is shown in Fig.~2(a). The
nonmonotonicity of $\Lambda$ as a function of $\alpha$ is typical, and
several dynamical transitions can be described: P $\to$ S, S $\to$ C,
and C $\to$ S (note that periodic motion here corresponds to a
n-frequency torus, with n = 1, 2 or 4; higher n such as 3, 6, 8 etc.
can also be observed in small windows of parameter space). 

The intermittency route to SNA that we describe in this work occurs
along the boundary of the chaotic region C$_2$ and the periodic region
T on its right, namely at the point marked I in Figs.~1 and 2a. An
enlarged view is shown in Fig.~2b where the bars indicate the variance
in the Lyapunov exponent estimated from several computations.  There is
a transition from a chaotic attractor to a SNA at $\alpha \approx
$3.405802\, with $\Lambda$ changing linearly \cite{L} through zero. The
{\it intermittent} transition from the SNA to a torus---the phenomenon
described here---occurs at $\alpha = \alpha_c \equiv $ 3.4058088$
\ldots$.  At this transition, the distinctive signature is an
abrupt change in the dependence of $\Lambda$ on the parameter $\alpha$,
as well as a concurrent marked reduction in the variance in $\Lambda$
(see Fig.~2b; the bars have been computed from 50 samples of $10^{5}$
steps).  The disappearance of the SNA or the chaotic attractor is
accompanied by intermittent dynamics, which can be conveniently studied
by co--evolving two trajectories with identical $(x_0,\phi_0)$ and
$\epsilon^{\prime}$, with different $\alpha$: since the angular
coordinate remains identical, the distance between the trajectories is
simply the difference in the $x_n$'s. We find that the time between
bursts shows the scaling $\tau \sim (\alpha_c -
\alpha )^{-\theta}$, with $\theta \approx 1/2$ (the numerical value we
obtain is 0.52$\pm$ 0.04). This transition to an intermittent SNA
(Fig.~2c) occurs all along the right boundary of C$_2$.  In the
$\alpha$-$\epsilon$ plane, this is a region of nearly constant
$\epsilon$.

The abrupt death of the strange nonchaotic attractor is through a
quasiperiodic analogue of the saddle--node bifurcation where a period 1
orbit is born. Consider a sequence of maps that are rational
approximations to Eq.~(\ref{HH}) by setting $\omega = \omega_k =
F_k/F_{k+1}$, where $F_k$ is the $k$th Fibonacci number. The
$F_{k+1}$th iterate of the map is a function of $x$ alone (since
$\phi_{F_{k+1}} \equiv \phi_0$). This allows for the construction of a
bifurcation diagram as a function of $\alpha$ for fixed $\epsilon$. The
case of $k = 2$, namely $\omega = 1/2$ has been studied earlier by
Sanju and Varma \cite{SV} as the period--2 modulated logistic map.
Between the first period doubling bifurcation and the 2--band merging
crisis, the attractor of the (unmodulated) logistic map consists of two
branches. (For the logistic map, these bifurcations occur at $\alpha =
3$ and $\alpha \approx 3.678857 \ldots$, respectively).  The effect of
period--2 modulation is to split the two branches of the attractor, by
shifting one branch to higher values of $\alpha$ \cite{SV}. The extent
to which the two branches are shifted relative to one another depends
on the value of $\epsilon$. The single attracting orbit of period--2 in
the unmodulated map becomes two separate attracting orbits for small
values of the modulation; and depending on how the branches are
shifted, namely on the strength of the modulation, two very different
attractors---for example chaotic and periodic---can coexist. For
sufficiently large $\epsilon$ the two branches can in fact become
disjoint, namely the periodic orbit of one branch of the attractor is
shifted beyond the 2--band merging point (see Fig.~2 of
Ref.~\cite{SV}). When this happens, the periodic attractor appears as
at a saddle--node bifurcation, giving rise to a single periodic
attracting orbit of period 1.

It can easily be verified that this mechanism operates with small
modifications for the higher rational approximations $\omega_k$; this
is a case of period--$F_{k+1}$ modulation \cite{SV}.  Thus, in the
limit of $\omega_k \to \omega = (\sqrt{5}-1)/2$, we suggest that the
disappearance of the chaotic attractor and the SNA is through this
analogue of a saddle--node bifurcation.

Viewed as a function of decreasing $\alpha$, this transition
shares some of the features of a widening crisis in unforced systems
\cite{GORY,PL,MR}. The discontinuous change of $\Lambda$ at a saddle--node
bifurcation is softened by the quasiperiodic forcing, and
for $\alpha<\alpha_c$, the Lyapunov
exponent shows the scaling (see Fig.~3)
\beq
\Lambda-\Lambda_c \sim (\alpha_c -
\alpha)^{\mu}, 
\eqn
with the exponent $\mu = 0.37\pm0.03$  at $\epsilon^{\prime} = 1$. To 
within the quoted error--bar, the same exponent is obtained when the
probability density in the burst phase is fit to a power law
\cite{PL,MR,GOY}. The remnants of the old torus which are now embedded in
the SNA constitute one of the weakly coupled components of the SNA.
These exponents stay approximately constant along the C$_2$ boundary.

There are at least two other mechanisms for the creation of SNAs that
are operative in this system. Along the boundary of the chaotic region
C$_1$ in Fig.~1, which appears at the end of the (truncated) period doubling
cascades, SNAs that are formed can appear as a result of the
collision between a period-doubled torus and its unstable parent
\cite{HH}. SNAs are also created through the process of fractalization
\cite{K} which occurs along the entire left edge of the region C$_2$
(for example the cases studied by Nishikawa and Kaneko \cite{NK}).
Along the C$_1$ boundary both mechanisms operate and representative
points are marked as B and A respectively in Fig.~2a. A fractalizing
torus gets increasingly wrinkled as parameters are varied, and
eventually goes chaotic, without any period-doubling bifurcation.
These two routes can be differentiated from the geometry of the SNA: in
the torus collision mechanism \cite{HH}, a period 2$^k$--torus gives
rise to a 2$^{k-1}$--band SNA, while through fractalization
\cite{K,NK}, the result is a 2$^k$--band SNA. 

The intermittent SNA is similar in many respects to a {\it reentrant}
phase: the SNA so created is structurally distinct from the SNA born
through other mechanisms. This distinction also extends to the
corresponding tori, and examination of successive iterates of the line
$(1/2,\phi)$ under the map $f$ of Eq.~(\ref{HH}) gives an indication of
how this occurs.  By an extension of the $\epsilon = 0$ case
\cite{CE}, this line will be attracted to the 1-torus,
\beq
\lim_{k \to \infty} f^k(1/2,\phi)  \to T. 
\eqn
(The attractor of the dynamics, whether periodic, strange nonchaotic or
chaotic, is confined in a strip in configuration space, the upper
boundary of which is the line $f(1/2,\phi)$, and the lower boundary is
the {\it infimum} of $f^j(1/2,\phi), j \ge 2$). The convergence
behaviour of the line to the attractor, quantified through the phase
sensitivity index \cite{PF} shows that for smaller values of
nonlinearity and driving, the convergence is slower than for higher
nonlinearity. Indeed, the torus stabilized at large $\alpha,
\epsilon^{\prime}$ (in Fig.~1) is relatively smooth, as opposed to the
highly structured and fractalizing torus at smaller values of $\alpha$.

To further characterise this distinction, we examine the distribution
of finite--time Lyapunov exponents $P(\Lambda_N)$ in Fig.~4, namely the
Lyapunov exponent computed from a $N$-step orbit on the attractor.
Although $\Lambda$ is negative, for short times the Lyapunov exponent
on a SNA is positive \cite{PF}. The distribution $P(\Lambda_N)$ has not
been studied in detail, but it is known to have a significant positive
tail which does not vanish even for large $N$.  For the intermittent
SNA, $P(\Lambda_N)$ decays slowly both as a function of the (local)
$\Lambda_N$ as well as $N$. This distinction is a consequence of the
fact that the variance in the $\Lambda$ increases drastically through
this transition (in Fig.~2b). In contrast, at the other transitions
from a torus to a SNA (or vice-versa), the variance does not change
significantly and with increasing $N$, the $P(\Lambda_N)$ becomes
essentially gaussian.  The difference in the distributions is not
visible for $N \sim 50$ (Fig.~4a) while at $N \sim 1000$ the distinct
character of the intermittent SNA is clearly seen (Fig.~4b).  These
observations are consistent with a coarser interweaving of the positive
and negative local Lyapunov exponent regions on the intermittent SNA.
Furthermore, this transition is robust to low-amplitude noise
\cite{noise}, as has been verified by adding a small random component
to the $x$ dynamics.  The scaling behaviour in $\Lambda$ and the
probability density in the burst phase were observed to hold \cite{MPR}
with slightly modified values of the exponent.

Such quasiperiodically driven systems are interesting from a variety of
points of view. A potential novel use of SNAs that can be envisaged is
for secure communications \cite{KO}: two independent SNAs can be easily
synchronized since the Lyapunov exponents are negative, while the
signals from such systems provide effective masking since they appear
erratic owing to the underlying strange geometry. 

In summary, in the present Letter we have described a new mechanism for
the creation of strange nonchaotic attractors through intermittency,
whereby a chaotic attractor is eventually replaced by a quasiperiodic
one. In the intermediate region, there are SNAs.  The distinctive
signature of this transition is a sharp change in the Lyapunov exponent
which shows large fluctuations and scaling behaviour on the SNA side of
the transition. This mechanism, which is operative at high nonlinearity
and large amplitude quasiperiodic forcing, proceeds via a {\it
stabilization} of chaotic motion and is thus like a reentrant phase
transition. 

The route to intermittent SNA is a general one. We have studied a
variety of related systems \cite{MPR} and find that the phenomenology
is very similar: the phase--diagram that is obtained for the driven
logistic map is typical.  Intermittent SNAs always occur in analogous
regions in parameter space, namely where a tongue of periodic behaviour
separates two chaotic areas with differing characteristics \cite{ISNA}.
We expect that in situations where the amplitude of quasiperiodic
forcing is an conveniently varied parameter for example, this mechanism
for the creation of SNAs may be experimentally observed.

This work was supported by grant SPS/MO-5/92 from the Department of
Science and Technology, India. We thank Sanju for discussions.

\begin{figure}
\label{Phase}
\caption{Phase diagram for the forced logistic map (schematic) obtained
by calculating the Lyapunov exponent as a function of $\alpha$ and the
rescaled parameter $\epsilon^{\prime}$ defined as $\epsilon^{\prime} =
\epsilon/ (4/ \alpha-1)$ in a $100 \times 100$ grid.  P and C
correspond to Periodic (torus) and Chaotic attractors. The shaded
region along the boundary of P and C corresponds to SNA (marked S).
The boundaries separating the different regions are convoluted, and
regions of SNA and Chaotic attractors are interwoven in a complicated
manner.  The intermittent SNA is found on the edge of the C$_2$ region
marked I.}
\end{figure}

\begin{figure}
\label{fig2}
\caption{(a) The largest Lyapunov exponent $\Lambda$ versus $\alpha$
at the upper edge of the phase plane, namely at $\epsilon^{\prime}=1$
and $2 \le \alpha \le 4$. C$_1$, C$_2$ correspond to the two chaotic
phases identified in Fig.~1. T denotes periodic motion (tori); SNAs are
found along the boundaries of C$_1$ and C$_2$, in the neighbourhood of
the points marked A, B and I.  (b) An enlarged view of Fig.~2a) near
the intermittent transition marked I.  The bars indicates the variance
in $\Lambda$ computed from 50 samples of length 10$^5$.  (c) The
Intermittent SNA at $\alpha =$ 3.405808 and $\epsilon^{\prime}$= 1. }
\end{figure}

\begin{figure}
\label{fig3}
\caption{Scaling of the probability density ($p_B$) in the burst phase
($\circ$) and $\Lambda$ ($\Box$) at the intermittent transition at I,
for $\alpha_c = 3.4058088\ldots$ and $\Lambda_c = -0.0283$ at
$\epsilon^{\prime} = 1$. The measured exponents are $\approx 0.37$.}
\end{figure}

\begin{figure}
\label{fig4}
\caption{(a) Distribution of $N$--step or local Lyapunov exponents,
computed from N = 50 step segments of a long trajectory for the
intermittent SNA at the point marked I contrasted with SNAs (born
through other mechanisms) at the points marked A and B in Fig.~2a.
These correspond respectively to I~($+$), A~($\nabla$) and B~($\Box$).
(b) Same as (a), except that N=1000.}
\end{figure}

\begin{references}
\bibitem{noise} Noise is added by a term of the form $\sigma \xi_n$,
where $\xi_n$ is a uniform random variable in the unit interval. At
$\epsilon^{\prime} = 1$, for instance, the transition survives for
additive noise of amplitude up to $\sigma = 10^{-6}$, while at lower
$\epsilon^{\prime}$ the transition is robust to even larger $\sigma$
$\approx 10^{-4}$. In other systems with intermittent SNAs the
transition may survive to higher noise levels.
\bibitem{ISNA} We have located intermittent SNAs in a number of
systems such as the forced sine--map, as well as higher dimensional
examples such as the ring and H\'enon maps \cite{SFKP} using this
strategy \cite{MPR}.
\end{references}
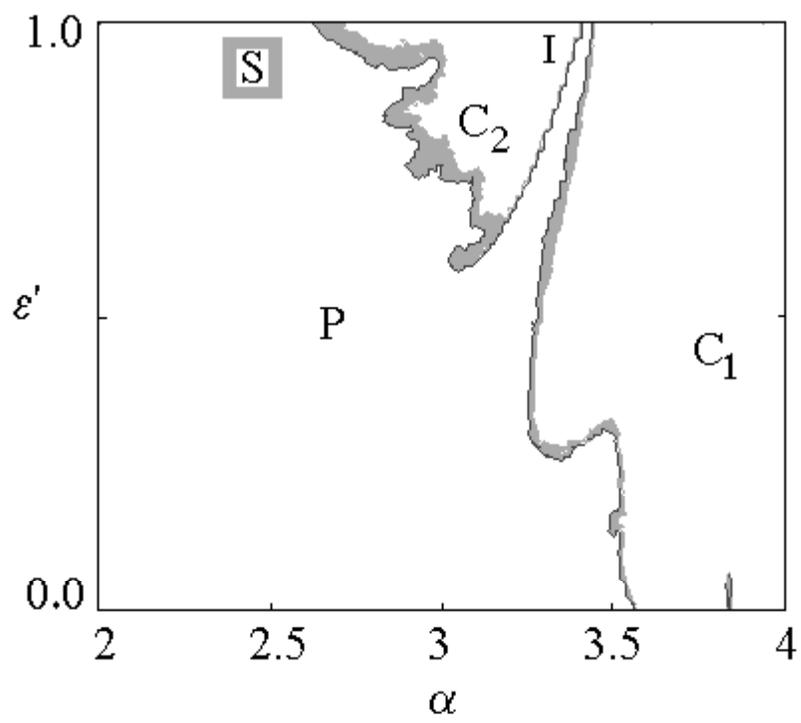
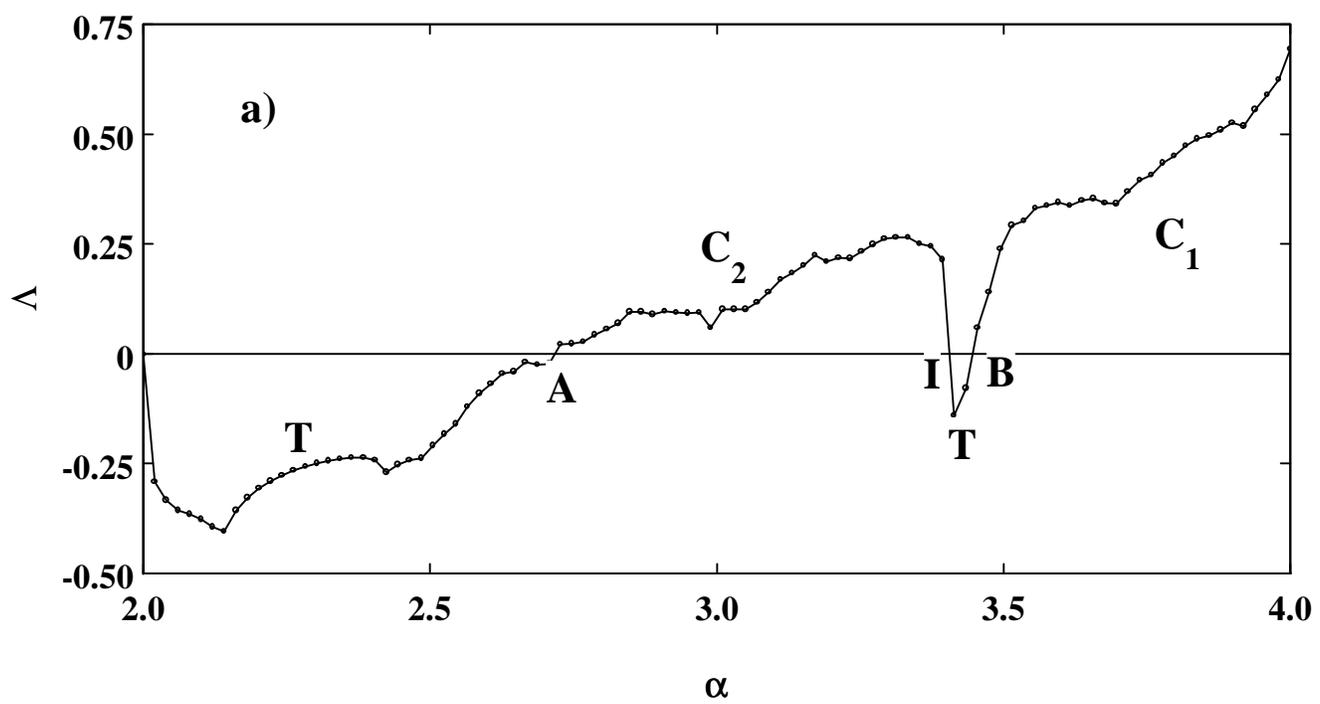
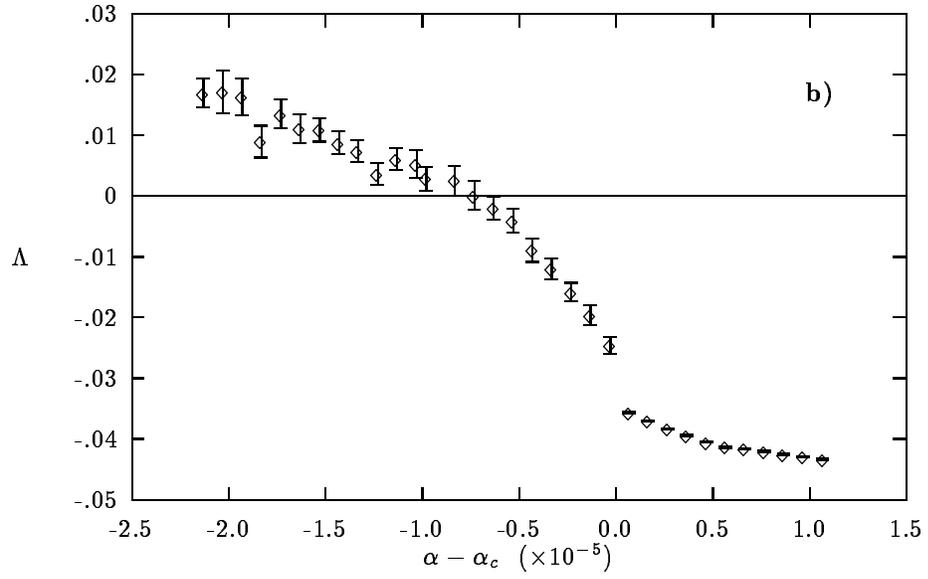
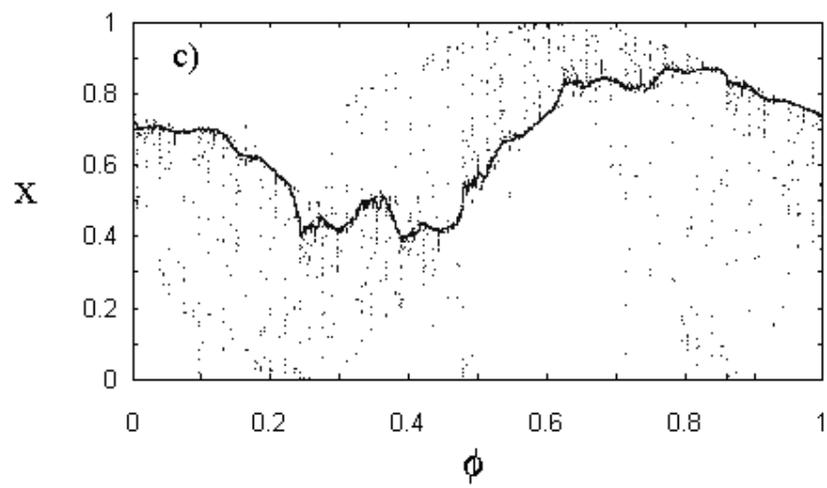
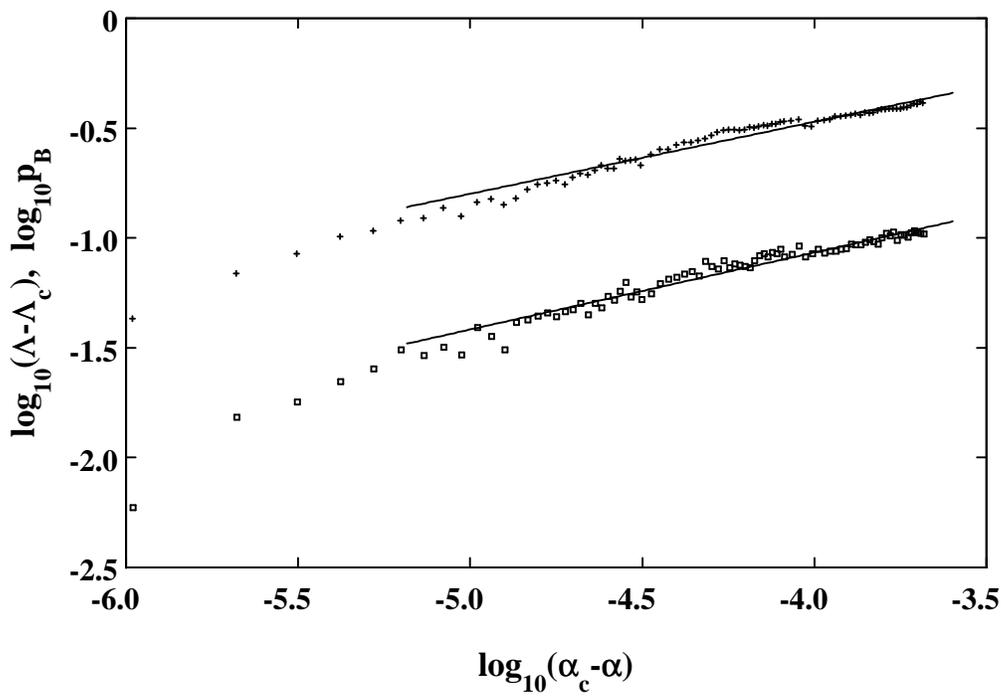
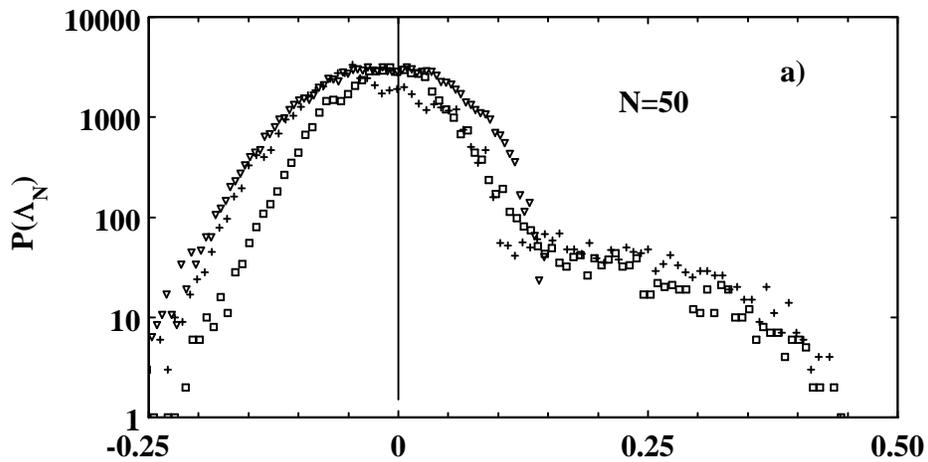
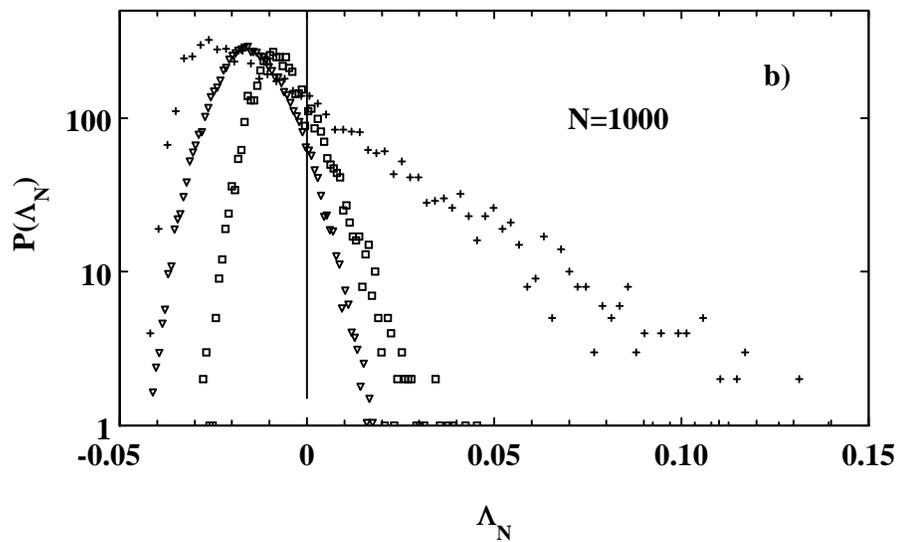
\end{document}